\documentclass[10pt,letterpaper,twocolumn]{article} 

\usepackage{OL}
\usepackage{hyperref}
\usepackage{amsmath}
\usepackage{amsfonts}

\usepackage{float} 

\begin{document}

\twocolumn[ 

\title{Manifestations of changes in entanglement and onset of synchronization in tomograms}

\author{Soumyabrata Paul,$^*$ S. Lakshmibala, V. Balakrishnan, and S. Ramanan}

\address{Department of Physics, Indian Institute of Technology Madras, Chennai 600036, India}

\email{$^*$Corresponding author: soumyabrata@physics.iitm.ac.in}


\begin{abstract}

Quantum state reconstruction for continuous-variable systems such as the radiation field poses challenges which arise primarily from the large dimensionality 
of the Hilbert space. Many proposals 
for state reconstruction 
exist, ranging from standard reconstruction protocols to applications of machine learning.  No universally 
applicable protocol  exists, however, for
 extracting the Wigner function from the optical tomogram of an arbitrary state of light.  We establish that nonclassical effects such as entanglement changes during dynamical evolution and the onset of quantum synchronization are mirrored in qualitative changes in optical tomograms themselves, circumventing the need for state reconstruction for this purpose.
\end{abstract} 


] 

\section{Introduction} \label{introduction}

It is now well known that speed-up in a wide range of computation and information processing protocols can be achieved in quantum platforms, as  compared to classical procedures. An important factor that is responsible for this advantage is the size of the state space which can be exploited for both storing and processing of information in quantum systems \cite{qst01, qst02, qst03}. 
As the dimension of the Hilbert space increases, however, quantum state reconstruction becomes increasingly 
formidable. For instance, reconstruction of new states of light, and tracking and characterization of the state of light at various instants of time in hybrid quantum platforms, still 
involve challenging open problems
\cite{lvovsky}. 
It is therefore worth avoiding, if possible, 
the  construction of the Wigner function from the raw data measured from experiments in the form of histograms. The focus would then be on extracting information about the state directly from tomograms. In this paper, we consider generic theoretical models of light interacting with an atomic medium, and establish that the onset of quantum synchronization\cite{saikat2020, bruder2018, bruder2020} and changes in the extent of quantum entanglement between subsystems can be identified directly from tomograms,\cite{manko1, sharmila2019, sharmila2020} avoiding the reconstruction program entirely. As a precursor to this, we point out that even in a single-mode system, at instants of revivals and fractional revivals of the wave packet of light,  qualitative changes can be  seen directly in tomograms\cite{sudheesh2015}. While these effects are also captured in the Wigner function, tomograms suffice for our purposes. This could prove potentially advantageous in state tomography.

The contents of this paper are arranged as follows. In Sec. \ref{tomograms} we discuss the salient features of optical tomograms and tomographic entanglement indicators. In Sec. \ref{Kerr} we consider a single-mode radiation field governed by the Kerr Hamiltonian, and illustrate how revivals and fractional revivals are reflected in the qualitative changes in appropriate tomograms. In Sec. \ref{AgarwalPuri}, using a bipartite model, we establish how tomograms capture changes in the extent of entanglement. For this purpose, we consider the interaction between a single-mode radiation field and a multi-level atom modelled as an oscillator. In Sec. \ref{lambda} we examine the onset of quantum synchronization and also changes in entanglement between the two fields in a tripartite hybrid quantum platform, and illustrate how reconstruction of the Wigner function is unnecessary for the identification of these phenomena. We conclude with a brief summary.

\section{Optical Tomograms and the Tomographic Entanglement Indicator} \label{tomograms}

A quorum of observables that  carries complete information about the state of  a single-mode radiation field 
is given by the set of rotated quadrature operators \cite{manko1}
\begin{equation} \label{tomogram:quad_op}
\mathbb{\hat{X}_{\theta}} = ( \hat{a}^{\dagger} e^{i \theta} + \hat{a} e^{- i \theta})/\sqrt{2},
\end{equation}
where $(\hat{a}^{\dagger}, \hat{a})$ are the photon creation and annihilation operators and $0 \le \theta < \pi$. The $x$ and $p$ quadratures correspond to $\theta=0$ and $\theta=\pi/2$, respectively.
The optical tomogram\cite{lvovsky} is
\begin{equation} \label{tomogram:w}
w(X_{\theta}, \theta) = \langle X_{\theta}, \theta | \hat{\rho} | \ X_{\theta}, \theta \rangle,
\end{equation}
where $\hat{\rho}$ is the density matrix and
\begin{equation}
\mathbb{\hat{X}_{\theta}} |X_{\theta}, \theta \rangle = X_{\theta} |X_{\theta}, \theta \rangle,
\end{equation}
with the normalization 
\begin{equation} \label{tomogram:int_w}
\int_{-\infty}^{\infty} \!dX_{\theta}~w(X_{\theta}, \theta) = 1
\end{equation}
for every value of $\theta$. 
For pure states, it is computationally advantageous to expand $w(X_{\theta}, \theta)$ in the photon number basis\cite{manko2}, in terms of
Hermite polynomials. The tomogram is plotted with $X_{\theta}$ as the abcissa and  $\theta$ 
as the ordinate. 
Generalizing to a bipartite system, the two-mode tomogram is
\begin{align} \label{tomogram:w_AB}
&w(X_{\theta_{A}}, \theta_{A}; X_{\theta_{B}}, \theta_{B}) 
\nonumber \\
& = 
\langle X_{\theta_{A}}, \theta_{A}; X_{\theta_{B}}, \theta_{B} | \hat{\rho}_{AB} |X_{\theta_{A}}, \theta_{A}; X_{\theta_{B}}, \theta_{B} \rangle
\end{align}
where $A, B$ label the two subsystems, and $X_{\theta_{A}}, X_{\theta_{B}}$ are the respective eigenvalues of
\begin{equation} 
\label{tomogram:two_mode_quad_ops}
\mathbb{\hat{X}_{\theta_{A}}} =  
\frac{
( \hat{a}^{\dagger} e^{i \theta_{A}} + \hat{a} e^{-i \theta_{A}})}{\sqrt{2}},
\mathbb{\hat{X}_{\theta_{B}}} =  
\frac{( \hat{b}^{\dagger} e^{i \theta_{B}} + \hat{b} e^{-i \theta_{B}})}{\sqrt{2}}.
\end{equation}
The reduced tomogram corresponding to $A$, 
 for instance, is given by
\begin{equation} \label{tomogram:reduced_tomogram}
w(X_{\theta_{A}}, \theta_{A}) = \int_{-\infty}^{\infty} \!dX_{\theta_{B}}~w(X_{\theta_{A}}, \theta_{A}; X_{\theta_{B}}, \theta_{B}) 
\end{equation}
for any given  $\theta_{B}$. 

The  inverse participation ratio (IPR), which quantifies the delocalisation of a state in a given basis,\cite{viola-brown} 
 is directly obtained from the tomogram. The two-mode IPR is defined as
\begin{align} \label{tomogram:eta_IPR}
\eta_{AB}(\theta_{A}, \theta_{B}) 
=  &\int_{-\infty}^{\infty} \!dX_{\theta_{A}} \int_{-\infty}^{\infty} \!dX_{\theta_{B}}  \times  \nonumber \\
& \left[ w(X_{\theta_{A}}, \theta_{A}; X_{\theta_{B}}, \theta_{B}) \right]^{2}.
\end{align}
The reduced subsystem IPRs are given by 
\begin{equation} \label{tomogram:eta_IPR_A}
\eta_{k}(\theta_{k}) = \int_{-\infty}^{\infty} \!dX_{\theta_{k}} \left[ w_{k}(X_{\theta_{k}}, \theta_{k}) \right]^{2} ~ (k = A, B).
\end{equation}
A useful tomographic entanglement indicator (TEI) 
that can be obtained 
from the IPR 
is then given by\cite{sharmila2019, sharmila2020} 
\begin{equation} \label{tomogram:epsilon_IPR}
\epsilon_{\text{IPR}}(\theta_{A}, \theta_{B}) = 1 - \big[\eta_{A}(\theta_{A}) + \eta_{B}(\theta_{B}) - \eta_{AB}(\theta_{A}, \theta_{B})\big].
\end{equation}
The dependence on $(\theta_{A}, \theta_{B})$ is removed by taking an average over the quorum, to get
\begin{equation} \label{tomogram:xi_IPR}
\xi_{\text{IPR}} \equiv \langle \epsilon_{\text{IPR}}(\theta_{A}, \theta_{B}) \rangle.
\end{equation}
We have verified that, in practice, 
 $25$ different combinations of $(\theta_{A}, \theta_{B})$ equispaced in $[0, \pi)$ 
for each $\theta_{i}$,  are adequate for the purpose 
at hand.

In the next section we establish that tomograms also suffice for the identification of revivals and fractional revivals that occur during the propagation of a radiation field in a nonlinear optical medium.

\section{Single-mode field propagating in a Kerr medium} \label{Kerr}

The effective Hamiltonian in this case \cite{robinett, Agarwal1993} is given by $\mathcal{\hat{H}}_{1} = \lambda \hat{a}^{\dagger 2} \hat{a}^{2}$,
where $\lambda$ is the third-order nonlinear susceptibility. (Here, and in the subsequent sections, we set $\hbar = 1$.) Consider an initial field coherent state (CS),
 given in the photon number basis \{$|k\rangle$\} by 
 $|\alpha\rangle = e^{-|\alpha|^{2}/2}\sum_{k=0}^{\infty} \alpha^{k}|k\rangle/
\sqrt{k!}$ where $\alpha \in \mathbb{C}$.
The state at any time $t$ is then given by 
$e^{-|\alpha|^{2}/2} \sum_{k=0}^{\infty}  \alpha^{k}
 e^{-i \lambda k(k - 1) t} |k\rangle/\sqrt{k!}$.
At instants that are integer multiples of the revival time $T_{\text{rev}} \equiv \pi/\lambda$, the field revives and the fidelity $|\langle \psi(0) | \psi(T_{\text{rev}}) \rangle|^{2} = 1$. Further, $p$-subpacket fractional revivals occur at integer multiples of $T_{\text{rev}}/p$ ($p = 2, 3, \ldots$), when the state is a superposition of $p$ coherent states. These  revival phenomena arise due to the periodicity properties of the unitary time evolution operator.

\begin{figure}[t]
\begin{center}
\includegraphics[width=0.5\textwidth]{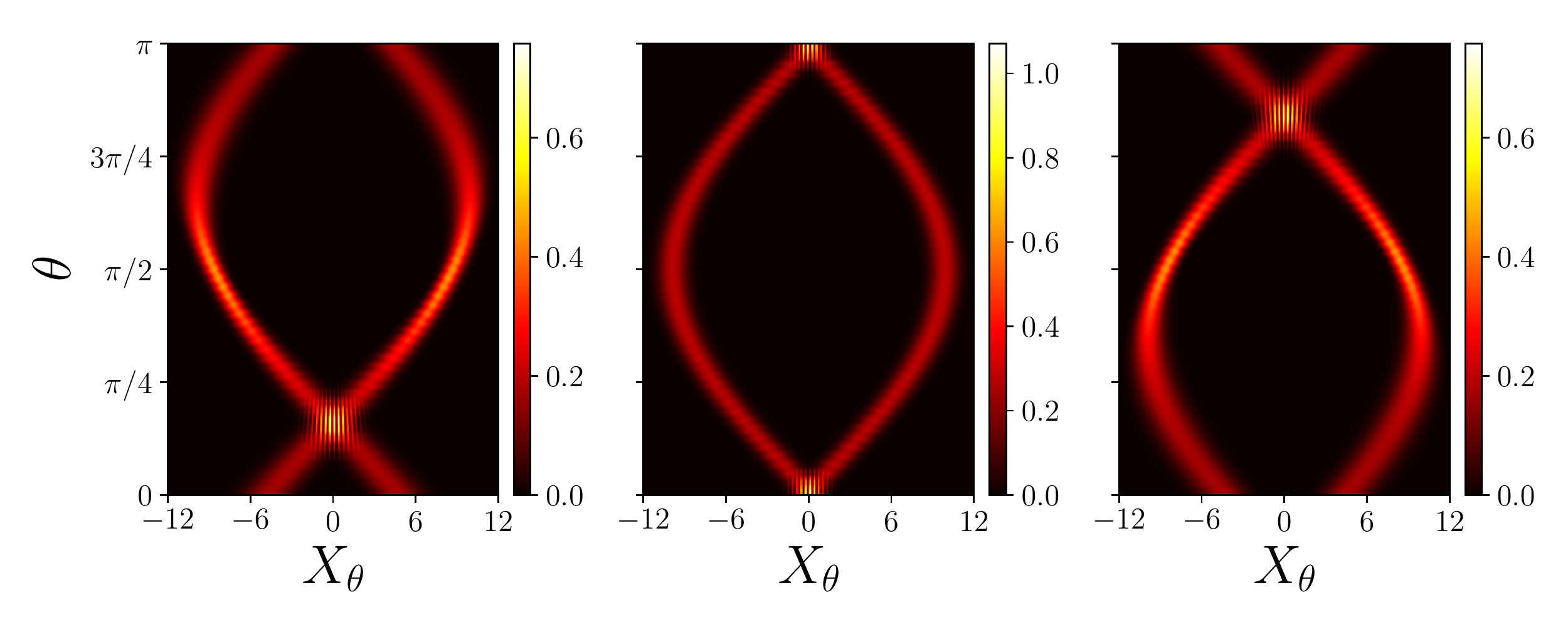}
\vspace{-5ex}
\caption{Left to right: Optical tomogram $w(X_{\theta}, \theta)$ for an initial CS $|\alpha\rangle$ propagating in a Kerr medium, at instants $t = \frac{1}{2} T_{\text{rev}} - \epsilon$, 
$t = \frac{1}{2} T_{\text{rev}}$, 
and $t = \frac{1}{2} T_{\text{rev}} + \epsilon$, for  
$|\alpha|^2=49$, $\lambda=1$, $\epsilon = 0.005$.}
\label{Kerr:fig_tomogram_Kerr_abs2_49_panel}
\end{center}
\end{figure}

Figure  \ref{Kerr:fig_tomogram_Kerr_abs2_49_panel} 
shows the clearly discernible   change in the tomogram at $\frac{1}{2}T_{\text{rev}}$ (the instant of two-subpacket fractional revival), compared to its appearance just before and just after 
 this instant. Such qualitative differences can be seen at all instants of fractional revivals and full revivals.

\section{Entanglement in a bipartite system} \label{AgarwalPuri}

\begin{figure}[t]
\begin{center}
\includegraphics[scale=0.35]{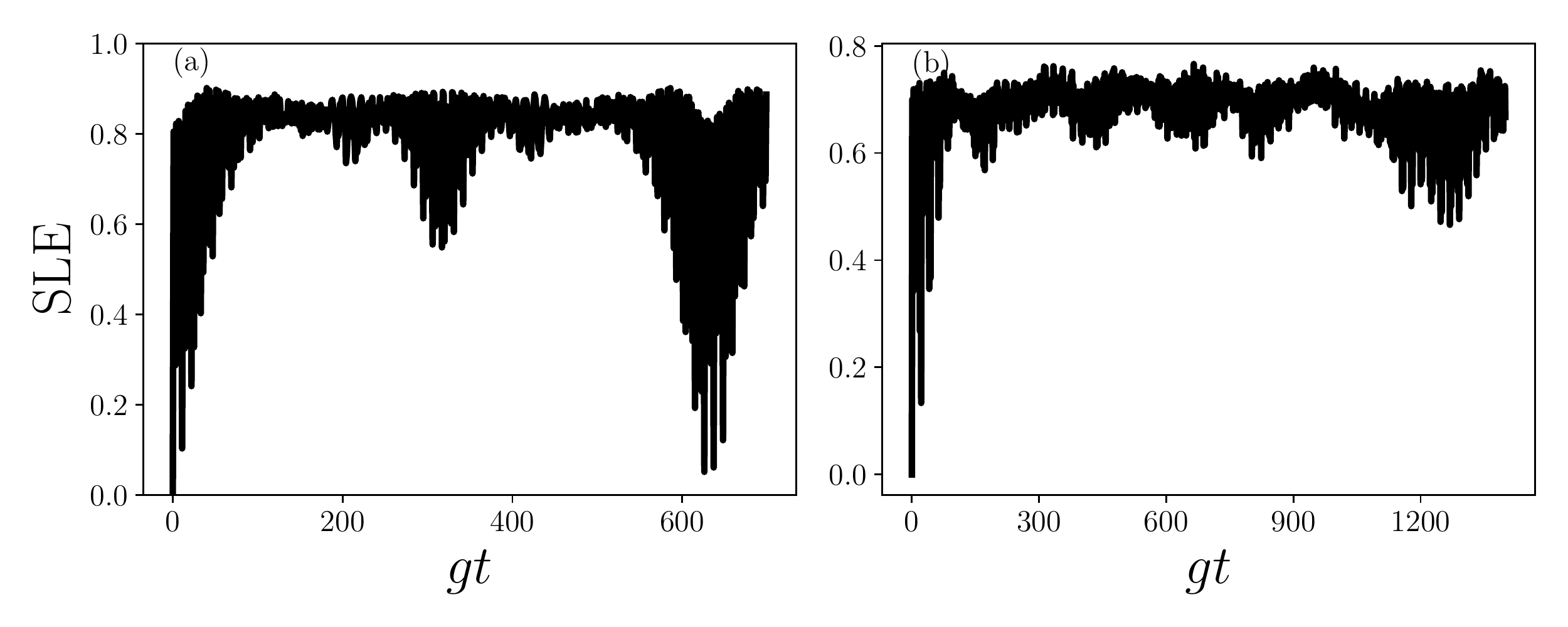}
\vspace{-5ex}
\caption{$\text{SLE}$ vs. $gt$ for an initial state (a) $|10; 0\rangle$  and (b) $|\alpha; 0\rangle$  
with $\alpha = \sqrt{5}$. Figure reproduced from Ref. \citeonline{sudheesh2006}.}
\label{AgarwalPuri:fig_sle_agarwal_puri_panel}
\end{center}
\end{figure}

The Hamiltonian for a bipartite model of a field with photon creation and annihilation operators $(\hat{a}^{\dagger}, \hat{a})$ interacting with a multilevel atom modelled as an oscillator with ladder operators $(\hat{b}^{\dagger}, \hat{b})$  is given by
\begin{equation} \label{AgarwalPuri:eq_H}
\mathcal{\hat{H}}_{2} = \omega \hat{a}^{\dagger} \hat{a} + \omega_{0} \hat{b}^{\dagger} \hat{b} + \gamma \hat{b}^{\dagger 2} \hat{b}^{2} + g ( \hat{a}^{\dagger} \hat{b} + \hat{a} \hat{b}^{\dagger} ).
\end{equation}
Here, $\omega$ and $\omega_{0}$ are the field and atomic frequencies,  $\gamma$ is the  nonlinearity strength, and $g$ is the field-atom coupling constant. The detailed dynamics of the system can be solved\cite{AgarwalPuri} recognising that $\hat{\mathcal{N}} = \hat{a}^{\dagger}\hat{a} + \hat{b}^{\dagger}\hat{b}$ is a constant of the motion, and that the eigenstates of $\mathcal{\hat{H}}_{2}$ are also eigenstates of $\hat{\mathcal{N}}$ with eigenvalues $N 
( = 0, 1, 2, \ldots)$. It is therefore convenient to diagonalize $\mathcal{\hat{H}}_{2}$ in the basis \{$|N - m\rangle_{a} \otimes |m \rangle_{b}\} \equiv \{|N - m; m \rangle$\}, where $m$ is the atomic level that  takes values $0, 1, \ldots, N$ for a given  $N$.

\begin{figure}[t]
\begin{center}
\includegraphics[scale=0.35]{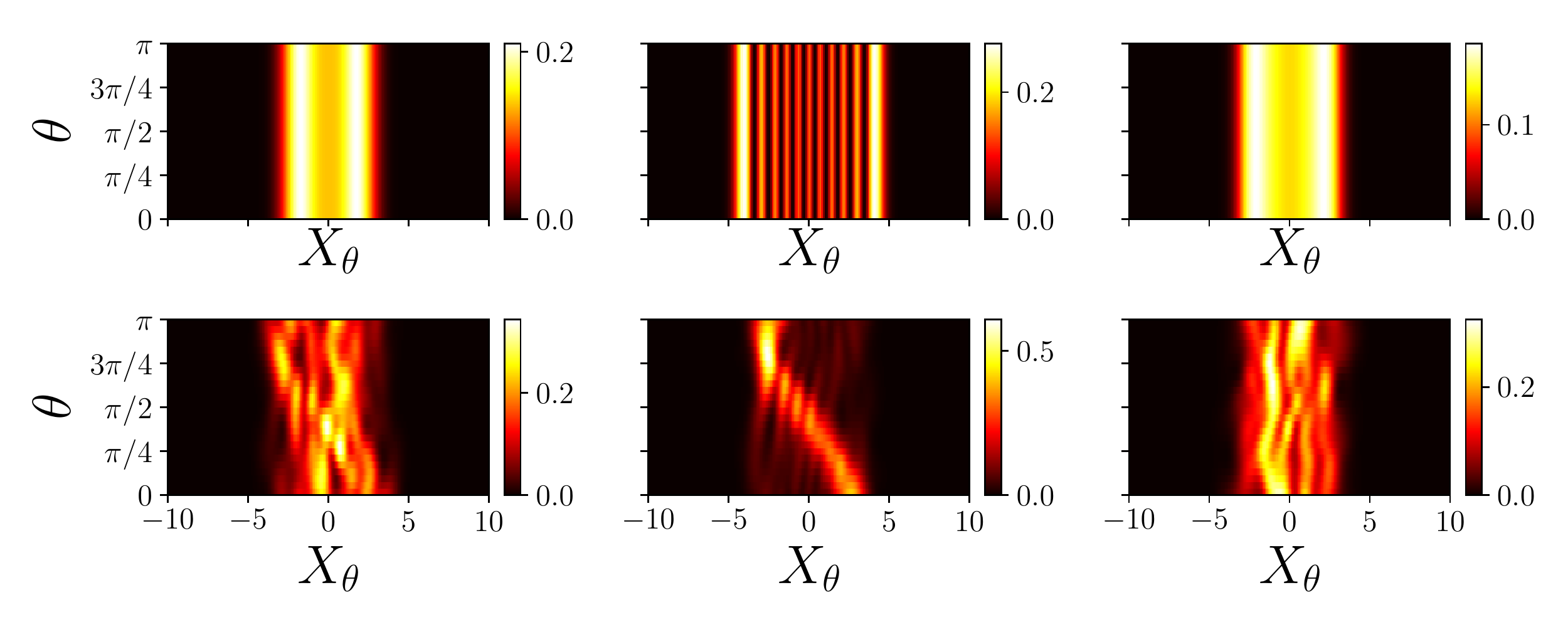}
\vspace{-5ex}
\caption{Left to right: Optical tomograms $w(X_{\theta}, \theta)$, for an initial state $|10; 0\rangle$ at $gt = 625, 626$ and $627$ resp. (top panel) and $|\alpha=\sqrt{5}; 0\rangle$ at $gt= 1268, 1269$ and $1270$ resp.  (bottom panel).}
\label{AgarwalPuri:fig_tomogram_rho_A_Fock_10_0_CS_sqrt_5_0_panel}
\end{center}
\end{figure}

The manner is which the bipartite entanglement varies with scaled time $gt$, for initial field states which are either Fock states or CS, has been examined elsewhere\cite{sudheesh2006} using the subsystem linear entropy (SLE) as a measure of entanglement. For ready reference, some salient results  are reproduced in Fig. \ref{AgarwalPuri:fig_sle_agarwal_puri_panel}. Here 
$\omega=\omega_{0}=1, \gamma/g=0.01$, the atom is initially in the ground state $|0\rangle$,  and the field is in (a) 
the Fock state $|10\rangle$,  (b) the CS $\alpha=\sqrt{5}$. In terms of $\rho_{a}$ (the reduced density matrix corresponding to the field) obtained by tracing out the atomic states from the full density matrix, $\text{SLE}(t) = 1 -\text{Tr} [\hat{\mathcal{\rho}}_{a}^{2}(t)]$.
From detailed numerical computations
it  has been reported earlier\cite{sudheesh2006} that dips in the SLE occur at specific instants of time, with the entanglement 
becoming very nearly equal to zero at $gt \approx 626$ (see Fig. \ref{AgarwalPuri:fig_sle_agarwal_puri_panel}(a)). 
The present investigation shows, however,   that these changes in the degree of entanglement can be identified readily from the qualitative changes in the field tomograms  in the neighborhood 
of  these instants, without any detailed 
calculations. We emphasize 
 that these features can be inferred 
 from the tomograms, totally avoiding  state reconstruction.
For illustrative purposes we consider the minimum at $gt = 626 ~ (\approx 200\pi)$, obtained numerically. Figure \ref{AgarwalPuri:fig_tomogram_rho_A_Fock_10_0_CS_sqrt_5_0_panel} (top panel) (tomograms at $gt=625, 626, 627$) demonstrates this observation. More dramatic qualitative changes in the tomogram arise if the initial field state is a CS. This is shown in Fig. \ref{AgarwalPuri:fig_tomogram_rho_A_Fock_10_0_CS_sqrt_5_0_panel} (bottom panel) by comparing tomograms at $gt = 1268, 1269,$ and $1270$. This feature is evinced  in the dip in 
the SLE at $gt = 1269$ in Fig. \ref{AgarwalPuri:fig_sle_agarwal_puri_panel}(b).

In the next section, we examine a tripartite hybrid quantum system and establish that both the onset of quantum synchronization and changes in the degree of entanglement are 
reflected  in tomograms.

\section{$\Lambda$-atom interacting with a bipartite radiation field} \label{lambda}

We now consider a $\Lambda$-atom with atomic levels $|e_{1}\rangle$, $|e_{2}\rangle$  and $|e_{3}\rangle$ interacting with two radiation fields $F_{i} \,(i = 1,2)$, with corresponding photon creation and annihilation operators $(\hat{a}_{i}^{\dagger}, \hat{a}_{i})$ and field frequencies $\Omega_{i}$. The fields 
  mediate the $|e_{i}\rangle \leftrightarrow |e_{3}\rangle$ transitions. Direct  $|e_{1}\rangle \leftrightarrow |e_{2}\rangle$
 transitions are  dipole-forbidden. The full Hamiltonian is given by
\begin{equation} \label{lambda:H}
\mathcal{\hat{H}}_{3} = \sum_{k=1}^{3} \omega_{k} \hat{\sigma}_{kk} + \sum_{i=1}^{2} \big[ \Omega_{i} \hat{a}_{i}^{\dagger} \hat{a}_{i} + \chi \hat{a}_{i}^{\dagger 2} \hat{a}_{i}^{2} + \kappa \left( \hat{a}_{i} \hat{\sigma}_{3i} + \text{h.c.} \right) \big].
\end{equation}
Here, $\hat{\sigma}_{jk} = | e_{j} \rangle \langle e_{k} |$ ($j, k = 1, 2, 3$), \{$\omega_{k}$\} are positive constants, $\chi$ is the strength of the field nonlinearity, and $\kappa$ is the atom-field coupling strength. Each field is initially in a CS $|\alpha\rangle$, and the atom is in $|e_{1}\rangle$. The 
explicit expression\cite{pradip2016} 
for the full system state at a subsequent time $t$ 
is a linear superposition of $| 1; m; n \rangle$, 
$|2; m - 1; n + 1 \rangle$ and $| 3; m - 1; n \rangle$ 
where  $|k; m; n \rangle$ stands for the 
product state $|e_{k}\rangle \otimes 
|m\rangle \otimes |n\rangle$, and
$|m\rangle,   |n\rangle$ are photon Fock states.
 In  what follows, we establish  that the onset of quantum synchronization between the two fields, and changes in the extent of their entanglement, are exhibited  very clearly in the field tomograms.

Synchronization in classical systems is heralded by limit cycles. In quantum systems several definitions of synchronization have been proposed. For our purposes,  we examine synchronization between $F_{1}$ and $F_{2}$ in the tripartite model
at hand using the synchronization indicator\cite{mari2013}
\begin{equation} \label{quant-synch:mari-sync-err}
S_{c}(t) = \langle \left( \Delta \hat q_{\_}(t) \right)^{2} + \left( \Delta \hat p_{\_}(t) \right)^{2} \rangle^{-1}.
\end{equation}
Here, $\langle (\Delta \hat A)^{2} \rangle = \langle \hat{A}^{2} \rangle - \langle \hat{A} \rangle^{2}$ is the variance of an operator $\hat{A}$,  and
$\hat q_{\_}(t) = (\hat q_{1}(t) - \hat q_{2}(t))/\sqrt{2}$, 
$\hat p_{\_}(t) = (\hat p_{1}(t) - \hat p_{2}(t))/\sqrt{2}$ 
where
$\hat q_{i}$ and $\hat p_{i}$ as the quadrature operators corresponding to field $F_{i}$. 

\begin{figure}[t]
\begin{center}
\includegraphics[width=0.45\textwidth]{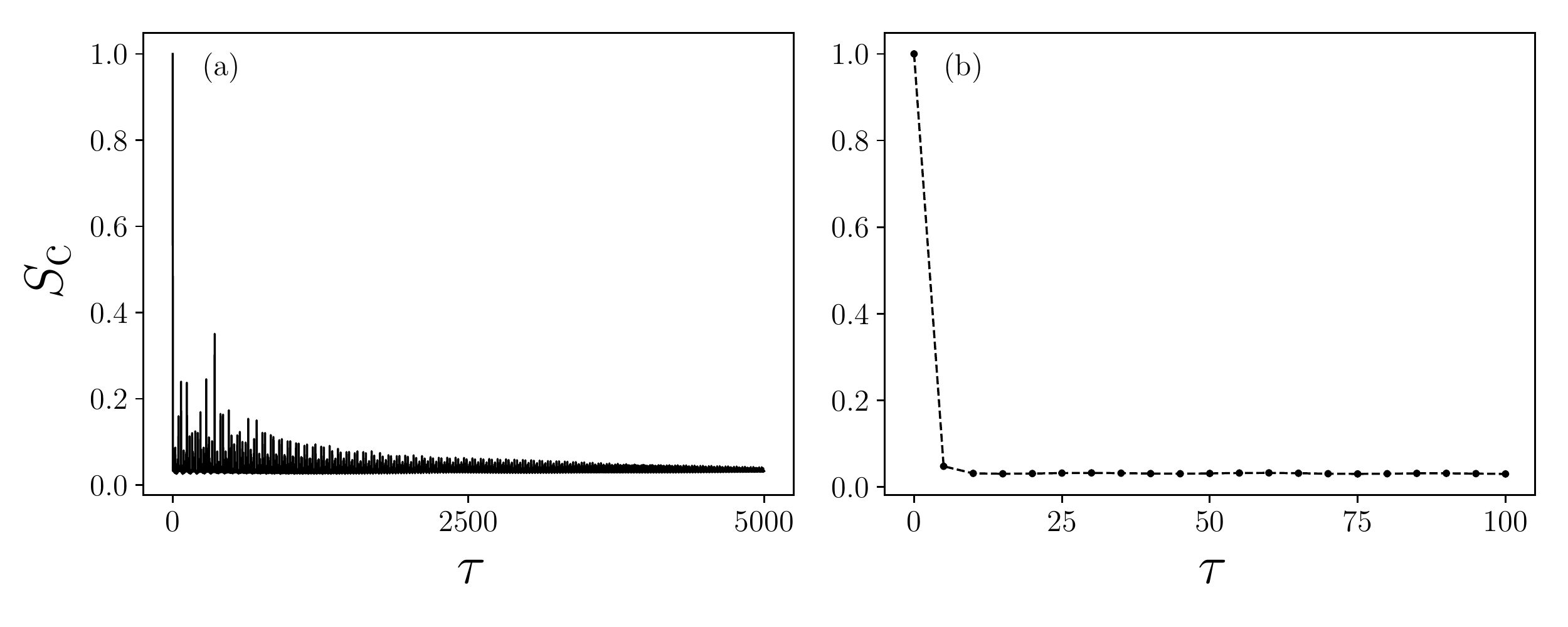}
\vspace{-3ex}
\caption{$S_{\text{c}}$ vs. $\tau$ for the $F_{1}$ subsystem from (a) $\tau = 0$ to $5000$ and (b) $\tau = 0$ to $100$, with $\Delta\tau=5$.}
\label{lambda:fig_S_c_panel}
\end{center}
\end{figure}

\begin{figure}[t]
\begin{center}
\includegraphics[width=0.45\textwidth]{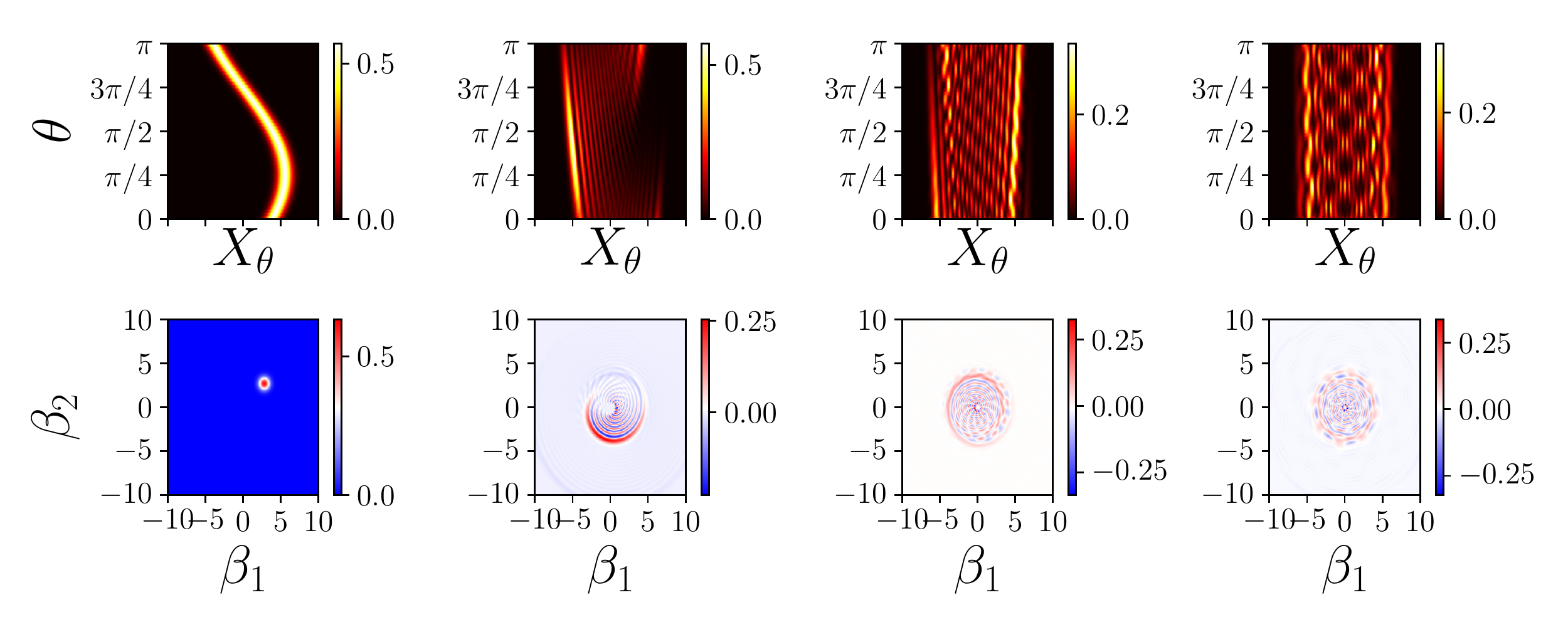}
\vspace{-3ex}
\caption{Optical tomogram $w(X_{\theta}, \theta)$ (top panel) and the corresponding Wigner function $W(\beta_{1}, \beta_{2})$ (bottom panel) for the field $F_{1}$ at instants $\tau = 0, 5, 10$ and $15$ respectively.}
\label{lambda:fig_lambda_rho_1_t_0_5_10_15}
\end{center}
\end{figure}

The parameters relevant to the dynamics can be shown to be $\kappa$, $\chi$, and $\Delta_{i} = \omega_{3} - \omega_{i} - \Omega_{i}$. In the  numerical computation, we have set  
$\kappa = 1$, $\chi = 5$, and $\Delta_{i} = 0$. Tracing out the atomic subsystem, a plot of $S_{c}$ versus  scaled time $\tau$ = $\kappa t$ reveals that the onset of synchronization between the two fields ($S_{c} \approx 0.03$) is at $\tau = 5$ (Figs. \ref{lambda:fig_S_c_panel}(a), (b)). This is revealed clearly in the qualitative change in the tomogram at this instant compared to the field tomograms at $\tau=0, 10, 15$ (Fig. \ref{lambda:fig_lambda_rho_1_t_0_5_10_15}, top panel). We have also computed the corresponding Wigner functions $W(\beta_{1}, \beta_{2})$, where 
$\beta_{1} = x/\sqrt{2}$ and $\beta_{2} = p/\sqrt{2}$ 
in terms of the field quadratures. 
While these Wigner functions (Fig. \ref{lambda:fig_lambda_rho_1_t_0_5_10_15},  bottom panel) do capture these changes, as is to be expected, we emphasize once again that the tomograms suffice to identify the onset of quantum synchronization, and can in fact be used for this purpose.

\begin{figure}[t]
\begin{center}
\includegraphics[width=0.45\textwidth]{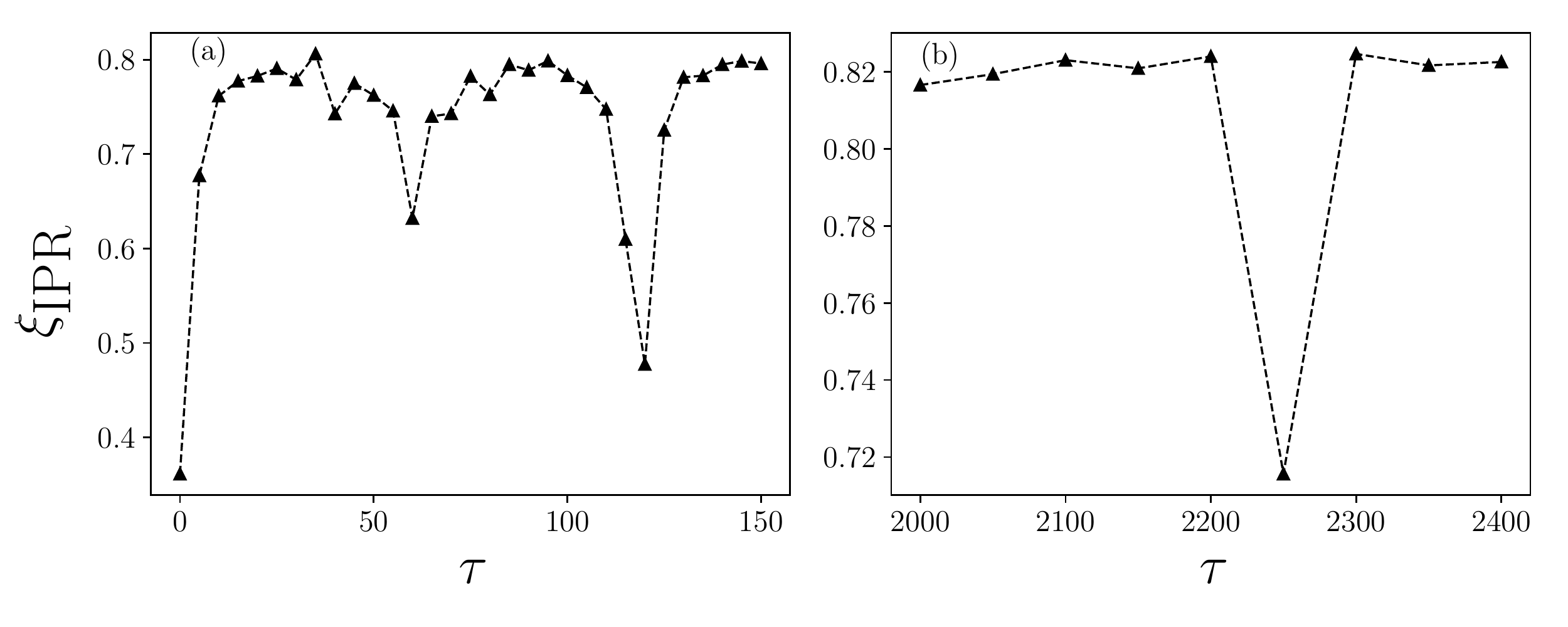}
\vspace{-3ex}
\caption{$\xi_{\text{IPR}}$ vs. $\tau$ for the bipartite field subsystem for (a) $\tau = 0$ to $\tau = 150$ with $\Delta\tau=5$, and (b) $\tau = 2000$ to $\tau = 2400$ with $\Delta\tau=50$.}
\label{lambda:fig_xi_IPR_panel}
\end{center}
\end{figure}

\begin{figure}[t]
\begin{center}
\includegraphics[width=0.45\textwidth]{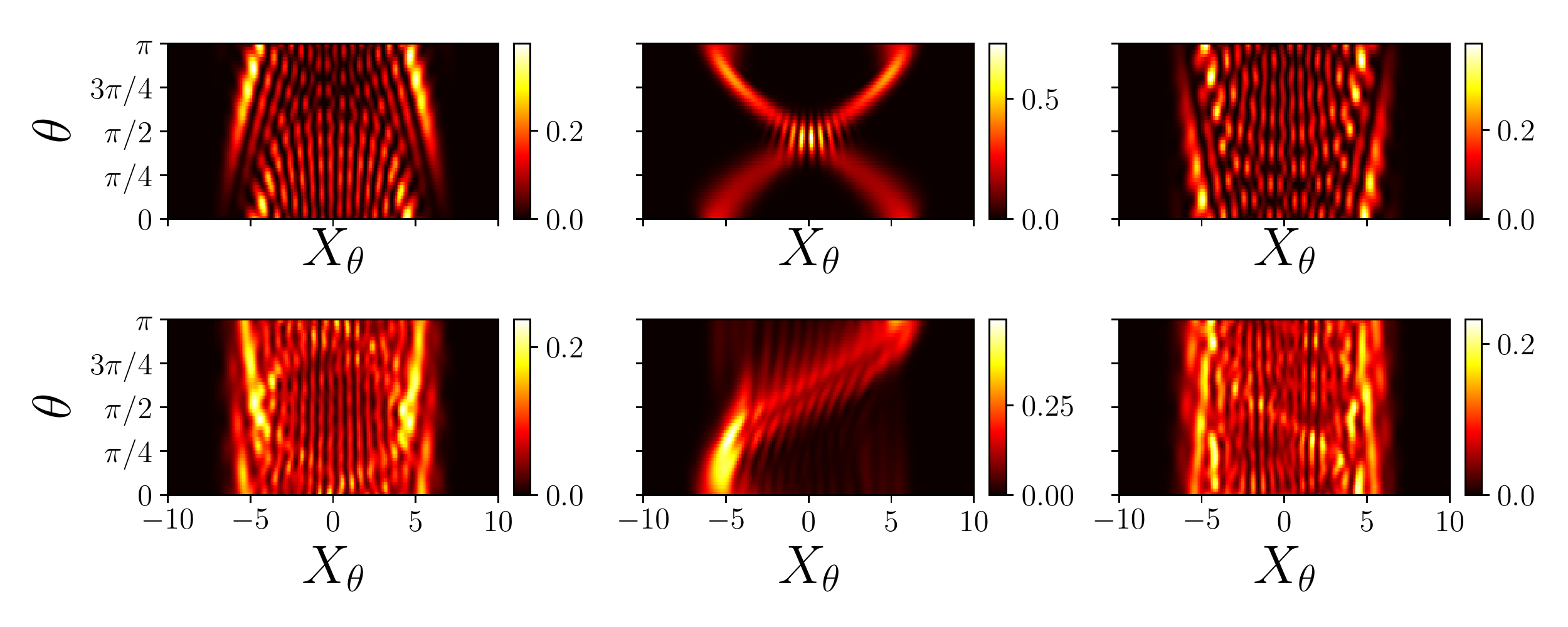}
\vspace{-3ex}
\caption{Left to right: Optical tomogram $w(X_{\theta}, \theta)$ at instants $\tau = 55, 60, 65$ resp. (top panel) and $\tau = 2200, 2250, 2300$ resp. (bottom panel).}
\label{lambda:fig_lambda_rho_1_t_60_2250_panel}
\end{center}
\end{figure}

Finally, we consider the  bipartite entanglement between the two fields.  The  plot of $\xi_{\text{IPR}}$ versus  $\tau$ in Fig. \ref{lambda:fig_xi_IPR_panel}(a) shows dips in the  entanglement at $\tau = 60$ and $120$. A comparison of the tomograms at $\tau = 55, 60,$ and $65$ (Fig. \ref{lambda:fig_lambda_rho_1_t_60_2250_panel},  top panel) captures this change in entanglement.  We draw attention to the fact that,  although the tomogram at $\tau = 60$ resembles single-mode tomograms at instants of two-subpacket fractional revivals, the former  is an entangled mixed state tomogram. Similar observations hold for 
the tomogram at $\tau = 120$. Again, in the long time dynamics, changes in entanglement are reflected in the corresponding changes in the  tomograms. For instance, the dip in $\xi_{\text{IPR}}$ at $\tau = 2250$ (Fig. \ref{lambda:fig_xi_IPR_panel}(b)) is captured in the tomograms in Fig. \ref{lambda:fig_lambda_rho_1_t_60_2250_panel} (bottom panel). 

In summary, we have established that optical tomograms (essentially histograms of 
appropriately chosen, experimentally measured quantities) suffice to identify the occurrence of revivals and fractional revivals in single-mode systems,  of  changes in bipartite entanglement in multipartite systems, and the onset of quantum synchronization between fields. Qualitative changes in the tomograms at these instants become more prominent as the number of subsystems increases. These observations therefore provide powerful pointers to identifying 
significant features in the dynamics of nonclassical effects, circumventing the need for detailed state reconstruction from tomograms.

\section*{Acknowledgment}
This work was supported in part by a seed grant from IIT Madras to the Centre for Quantum Information Theory of Matter and Spacetime, under the IoE-CoE scheme.

\end{document}